\documentclass[%
reprint,
superscriptaddress,
showkeys,
showpacs,preprintnumbers,
 amsmath,amssymb,
floatfix,
]{revtex4-2}

\usepackage[USenglish]{babel} 
\usepackage[T1]{fontenc}
\usepackage[ansinew]{inputenc}
\usepackage{tabularx}
\usepackage{lmodern} 
\usepackage{graphicx} %%For loading graphic files
\usepackage{amsmath}
\usepackage{amsthm}
\usepackage{amsfonts}
\usepackage{gensymb}
\usepackage{booktabs}
\usepackage{filecontents}
\usepackage{color}
\usepackage{mhchem}
\usepackage{upgreek}
\usepackage{xcolor}

\usepackage{mathtools}
\usepackage{booktabs} % Tarjoaa \toprule-, \midrule- ja \bottomrule-komennot taulukkoja varten
\usepackage{siunitx} 
\begin{document}

\title{Controlling thermal conductance using three-dimensional phononic crystals}

\author{Samuli Heiskanen}
\author{Tuomas Puurtinen}
\author{Ilari J. Maasilta} \email{maasilta@jyu.fi }
\affiliation{Nanoscience Center, Department of Physics, University of Jyvaskyla, P. O. Box 35, FIN-40014 Jyvaskyla, Finland}

\begin{abstract}
Controlling thermal transport at the nanoscale is vital for many applications. Previously, it has been shown that this control can be achieved with periodically nanostructured two-dimensional phononic crystals, for the case of suspended devices. Here we show that thermal conductance can also be controlled with three-dimensional phononic crystals, allowing the engineering of the thermal contact of more varied devices without the need of suspension in the future. We show experimental results measured at sub-Kelvin temperatures for two different period three-dimensional crystals, as well as for a bulk control structure. The results show that the conductance can be enhanced with the phononic crystal structures in our geometry. This result cannot be fully explained by the simplest theory taking into account the coherent modification of the phonon band structure, calculated with finite element method simulations.   
\end{abstract}


\maketitle

\section{Introduction}

Phononic crystals (PnC) are periodic structures which are elastic analogues of the more widely known photonic crystals \cite{joannopoulos2008molding}. Instead of a periodic dielectric constant, they have a periodic density and elasticity. This means that they can be used to manipulate how vibrational energy (sound and/or heat) flows through a material. The flow of energy can be altered due to changes introduced to the material's phononic band structure. PnC's can even produce full band gaps at specific frequencies due to Bragg interference \cite{PhysRevLett.71.2022, SIGALAS1993141}, or due to local resonances \cite{pennec2010two,New4}. Due to the ease of fabrication, early experiments focused on macroscopic structures with periods in the millimeter-scale \cite{PhysRevLett.93.024301}. Structures like this are typically applied in acoustic filtering, focusing or wave-guiding \cite{PhysRevLett.93.024301, tol20193d, pennec2010two}, because their dominant frequency is in the range of sonic or ultrasonic waves. Advancements in fabrication methods, like colloidal crystallization and interference lithography, allowed the fabrication of structures with micro- and nano-scale periods which have hypersonic dominant frequencies \cite{PhysRevLett.94.115501, cheng2006observation}. Such structures can be applied for example in RF communication devices \cite{olsson2008microfabricated}, but also for thermal transport, as hypersonic $\sim 10$ GHz frequencies are dominant for thermal phonons at low temperatures.  However, not much work has been done on thermal properties of such micro- and nano-scale PnC structures (for reviews, see \cite{New4, New1}). This is especially the case with low temperatures studies, where coherent modification of band-structure is still operational and not destroyed by disorder \cite{wagner,Mairee1700027}. There has been low temperature experimental studies on 2D holey PnC structures fabricated into a thin SiN film \cite{Zen_ncomms, Tian_2019}, but no such studies have been done on 3D PnCs before. Due to the difficulty of the fabrication of nano-scale 3D crystals, even some recent experiments on wave propagation have focused on macro-scale structures \cite{doi:10.1063/1.4971290}.

The control of thermal transport has become vital for several applications. There is a need to improve the heat dissipation out of semiconductor devices and better thermal isolation is needed for example in ultrasensitive bolometric radiation detectors \cite{Enss}. %\cite{wei2008ultrasensitive}.  
Similarly there are plans to use PnC structures to control thermal transport in quantum bits \cite{New3}. Previously this kind of control was achieved by introducing scattering centers, such as nanoparticles or impurities, into the material \cite{kim2006thermal, New5, lee2015thermal}. However, it has been shown that thermal transport can also be controlled with periodic PnC structures \cite{yu2010reduction, hopkins2011reduction, anufriev2016reduction}. It has been theoretically and experimentally demonstrated \cite{Zen_ncomms}, that at low temperatures PnCs alter the thermal conductance of the material by the coherent modification of its phononic band structure. This was shown for 2D crystals but similar ideas can be pursued also for 3D crystals, as discussed in this study. 
Considering different applications, the benefit of 3D crystals is that they make the isolation of any type of device possible, while 2D crystals only allow the isolation using more fragile suspended structures. 

It has been shown that self-assembly can be used for the fabrication of 3D PnCs with colloidal crystallization of mono-disperse spherical particles \cite{cheng2006observation, colloidalmultilayer, Isotalo_2012, Isotalo_2014}. However, there is no easy way to integrate any measurement devices and circuitry with such crystals. There have been efforts to allow fabrication of metallic wires on these crystals by hardening the spheres with an electron beam \cite{ Tian_2017}, but issues still exist with the continuity of the wiring due to the cracking of the crystals and separation at the substrate interface during drying. In this study, in contrast, 3D laser lithography is used for the direct crystal fabrication. The measurement devices and wiring were fabricated using a method developed for general device fabrication on high topographies, which utilizes the same 3D lithography tool, but in combination with lift-off \cite{samuli2}. The 3D lithography technique is based on two photon absorption, which allows fabrication of arbitrary 3D structures from negative resists \cite{sun,deubel,kawata}. The resolution of this technique is about 200 nm. Compared to colloidal crystallization, 3D lithography is the more versatile fabrication method, which also means that devices can be integrated more easily with the 3D structures. However, we are limited to micron-scale crystal periods due to the resolution limit of the technique. 

\begin{figure}
	\centering
    	\includegraphics[width=\columnwidth]{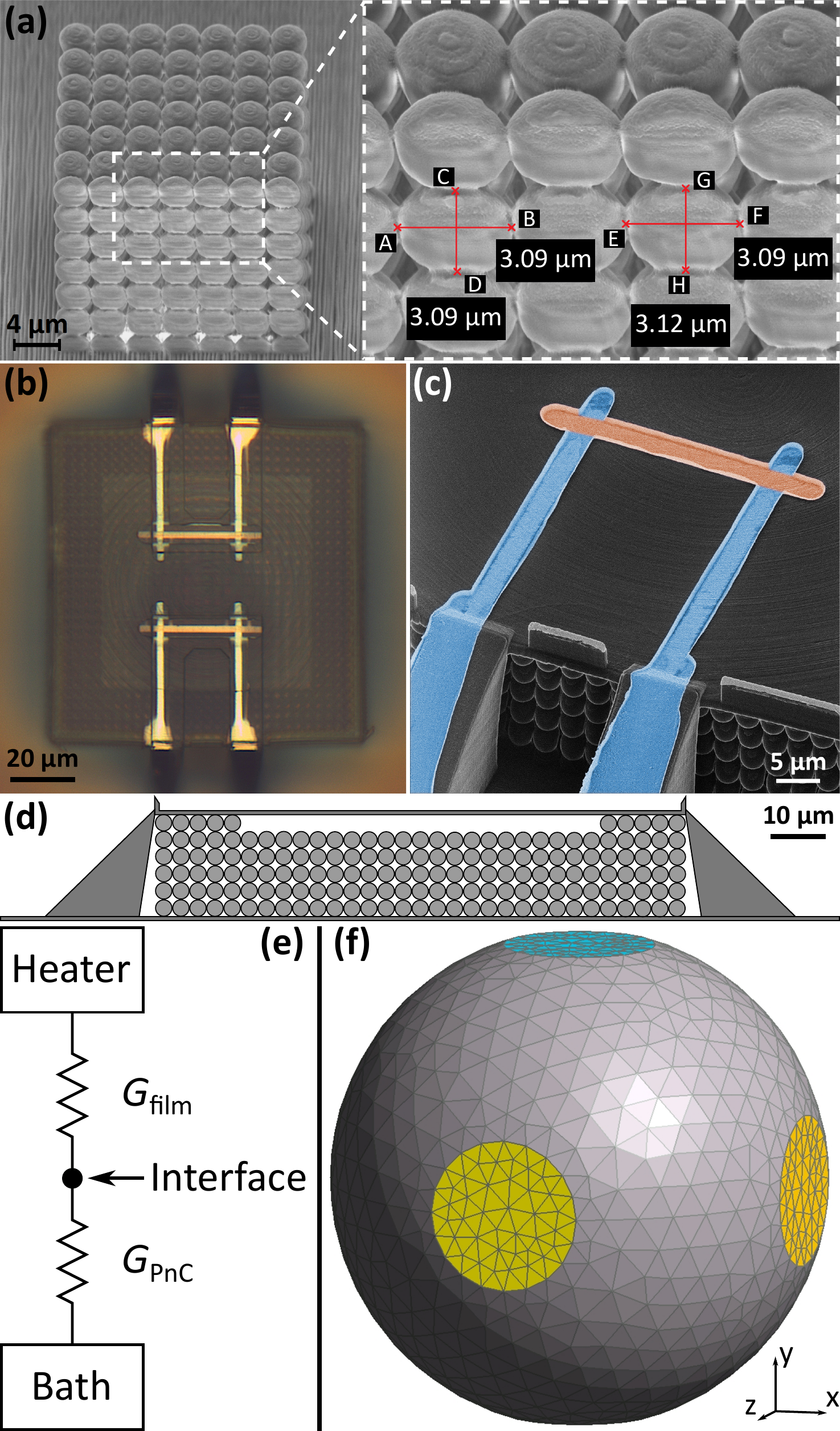}
    	\caption{(a) An SEM image of the 3.1 $\upmu$m PnC lattice with halved spheres at the front face. The image was taken at a 45\degree angle so the half spheres appear to be elliptical. (b) An optical microscope image of the finished structure from the top, including the heater and thermometer devices. (c) A false color helium ion micrograph of a finished SINIS junction pair on a PnC structure (blue=Al, red=Cu). The line width is around 3 $\upmu$m on the platform, giving a 9 $\upmu \text{m}^2$ junction area. (d) A schematic cross-section of the 3.1 $\upmu$m PnC structure.(e) The thermal model used in the analysis. (f)The FEM mesh used in the PnC simulations. For clarity, only the surface of the unit cell is shown, the mesh extends into the sphere volume. Contact areas between neighboring spheres are shown in yellow and blue.}
    	\label{fig:device}
\end{figure}

Here, we show with measurements and theoretical simulations that 3D PnC's fabricated with 3D lithography can %alter the coherent band structure of the material, and thus they can
 be used to control sub-Kelvin thermal transport by orders of magnitude. We fabricated 3D simple cubic lattices of spheres to obtain 3D PnCs that support a heated platform (Fig.~\ref{fig:device}). This geometry was chosen for simplicity of fabrication and modeling. Crystals with two different lattice constants were fabricated: One with a smaller lattice constant (3.1 $\upmu$m) that still resulted in a good spherical shape, and another larger period one (5.0 $\upmu$m), which could still be feasibly simulated using the finite element method (FEM). The size of the crystal and the filling factor of the lattice were the same for both PnCs, which means that classical diffusive bulk scattering would give identical thermal conductances for both structures.  However, our experiments show an order of magnitude difference between the two. Coherent simulations indeed predict a difference, but also a strong suppression compared to a bulk structure. Unexpectedly, for both structures, the conductance was enhanced compared to a bulk control device. %, which is in contradiction with our simulations. 
The reason for this is not fully clear, but is tentatively ascribed to an improved effective boundary conductance between the heated platform and the underlying structure. 

\section{Methods}

The PnC structures are fabricated with a Nanoscribe Photonic Professional 3D lithography system. The 3D exposure was done with the dip-in mode where an objective is brought in direct contact with a liquid resist. For the structures in this study we used the Ip-Dip resist by Nanoscribe GmbH. After the pattern of the structure is exposed into the liquid resist, development is done in propylene glycol methyl ether acetate (PGMEA) for 20 minutes and the sample is rinsed in isopropanol (IPA). Then a post-print UV curing is done for the samples \cite{Oakdale:16}, which strengthens the structure and minimizes the shrinkage of the resist, as follows: Keeping the sample in solution, it is directly transferred to an IPA bath containing 0.5 wt\% 2,2-dimethoxy-2-phenylacetophenone (DMPA) and then exposed to a 366 nm UV light for 20 minutes. The added DMPA acts as a photoinitiator, and thus this process increases cross-linking in the structure. Then the sample is rinsed again with IPA and dried with \ce{N_2} gas. The PnC structures are finalized by adding a 200 nm AlOx capping layer using a high vacuum electron-beam evaporator, in order to make the samples more durable for the following steps. This increased the yield considerably. The samples are baked on a hotplate at \SI{150}{\celsius} for 20 minutes before and after the AlOx evaporation in order to relieve internal stresses and reduce cracking.

%\begin{figure}
	%\centering
   % 	\includegraphics[width=0.7\columnwidth]{sphere_mesh.pdf}
    %	\caption{The FEM mesh used in the PnC simulations. For clarity, only the surface of the unit cell is shown, the mesh extends into the sphere volume. Contact areas between neighboring spheres are shown in yellow and blue.}
    %	\label{fig:mesh}
%\end{figure}

To facilitate the fabrication of a heater and a thermometer, the PnC structures have a smooth platform for them on top and ramps for measurement leads. The top layer of spheres is left out except at the edges of the platform so that it is suspended in the middle (Fig.~\ref{fig:device}d). The platform has a thickness 1.6 $\mu$m, which is well within the 3D limit for the thermally dominant phonon modes. This geometry was chosen to reduce the direct ballistic flow of heat from the heater to the crystal, which would not create a signal at the thermometer, thus the chose geometry makes sure that the thermometer signal is measurable. The heater and the thermometer are superconductor-insulator-normal metal-insulator-superconductor (SINIS) junction pairs fabricated opposing each other (Fig.~\ref{fig:device}b). The junction pairs are fabricated onto the middle of the platform using a fabrication method developed by us for device fabrication on 3D topographies \cite{samuli2}. For these samples aluminum was used as the superconductor and copper as the normal metal. The junction pairs can be used to measure the thermal conductance, by biasing one of them to work as a thermometer and then heating the film with the other junction pair \cite{Zen_ncomms}.

\begin{figure}
	\centering
    	\includegraphics[width=0.7\columnwidth]{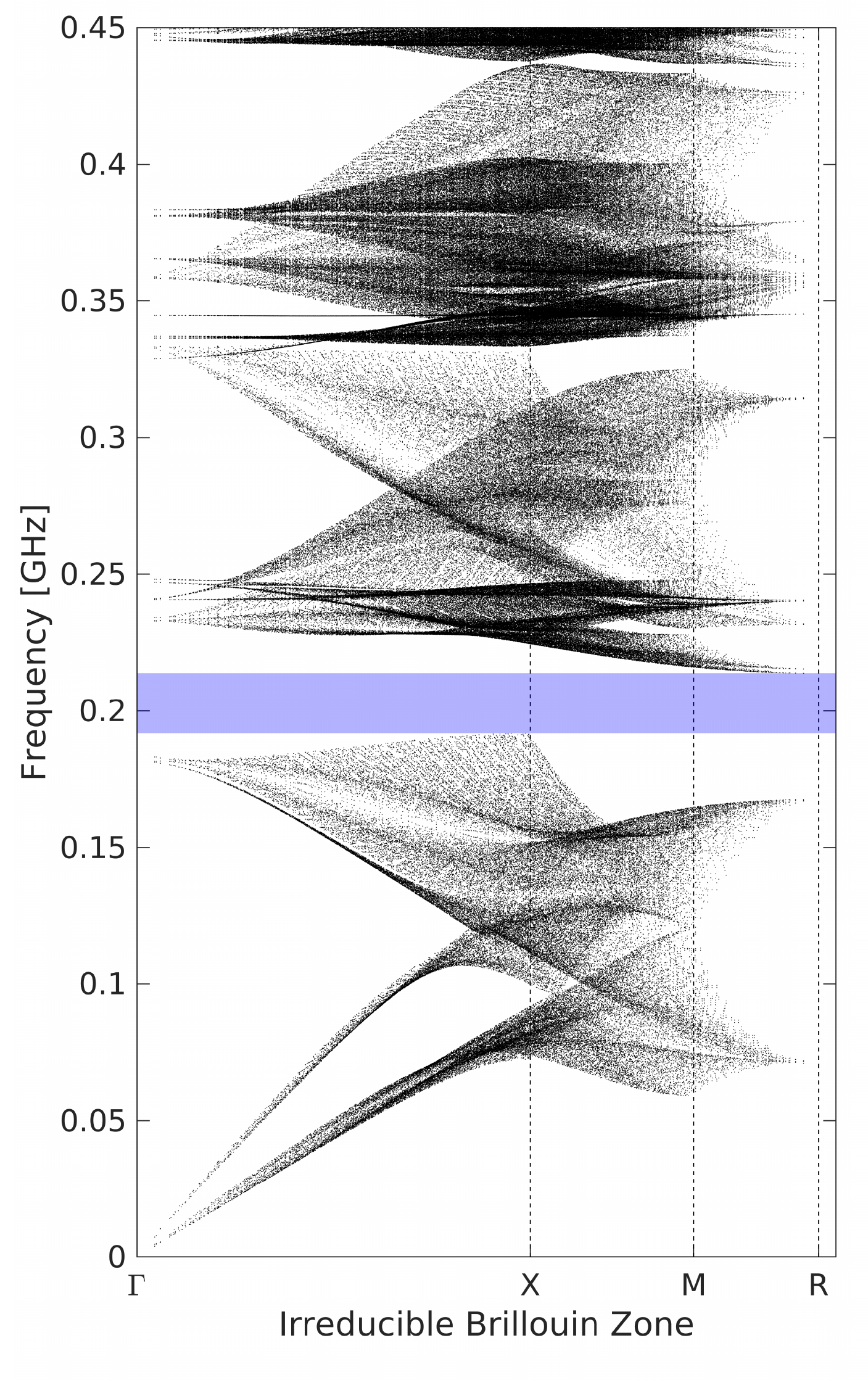}
    	\caption{The lowest spectral branches $\omega(\mathbf{k})$ given by the FEM simulations for the 3.1 $\upmu$m PnC. x-axis is the magnitude of the wave vector, with the distances to X,M, and R high symmetry points in the Brillouin zone noted. A complete energy gap is marked in blue.}
    	\label{fig:dispersion}
\end{figure}

The sample geometry could be approximated as a 3D sphere array covered by a film with a long rectangular heater on top of it. With the low temperatures ($<$1 K) used in the experiments, the flow of the phonons is typically ballistic \cite{Zen_ncomms,bolometers}. This means that the phonon emission is radiative and not diffusive. In the ballistic limit, the thermal conductance of the system can be calculated from the phonon emission power of the heater. The phonons emitted by the heater travel outwards in all directions inside the suspended platform. In turn, the area of the platform the overlaps with the PnC can be considered to be a large area phonon source ("heater") for the underlying PnC. In principle these two parts of the sample give two contributions in series (Fig.\ref{fig:device}e). However, with the used dimensions of the samples, we calculated (using Rayleigh-Lamb theory \cite{graff1975wave,PhysRevB.70.125425}) that  the thermal conductance of the platform is much higher than that of the PnC at the temperature range of the experiment 0.1 - 1 K, and does not limit the heat flow. The conductance of the platform is high, as it has essentially bulk behavior due to the softness of the material and the thickness of the platform (1.6 $\upmu$m), as the 2D-3D cross-over thickness $d_C=\hbar c_t/(2k_BT)$ \cite{Kuhn_2007} can be calculate to be as low as $\sim 30$ nm at 0.1 K for the IP-dip resist material.

With this information we can calculate the phonon emission power of the platform which flows through the underlying 3D PnC structure. Only the outward propagating phonon modes with energies $\hbar\omega_j(\mathbf{k})$ are carrying the energy, so the phonon emission power is given by the 3D version of the expression in  \cite{Puurtinen_crystals}:

 \begin{equation}
 \begin{split}
	\label{eq:heat_power}
	P(T)=&\frac{1}{8\pi^3}\sum_{j}\int_{\gamma}d\gamma\int_{K}d\mathbf{k}\hbar\omega_j(\mathbf{k})n(\omega_j,T) \\
	&\times\frac{\partial\omega_j(\mathbf{k})}{\partial\mathbf{k}}\cdot\hat{\mathbf{n}}_\gamma\Theta\left(\frac{\partial\omega_j}{\partial\mathbf{k}}\cdot
	\hat{\mathbf{n}}_\gamma\right),
\end{split}
\end{equation}

\noindent
where $\gamma$ represents a planar rectangular phonon source ("heater") element with area $A$, %length $l$ and width $d$.
 $\Theta$ is the Heaviside step function, $\hat{\mathbf{n}}_\gamma$ is the unit normal on the heater, and the k-integral is performed over the first 3D Brillouin zone $K$. Here $n(\omega,T)$ is the Bose-Einstein distribution describing the thermal occupation of the phonons and the term $\partial\omega_j/\partial\mathbf{k}$ describes the group velocity of the modes $j$. The only unknown in the expression is the set of dispersion relations $\omega_j=\omega_j(\mathbf{k})$, which can be calculated numerically using finite element modeling (FEM) of continuous linear elasticity theory \cite{Zen_ncomms}. Here we used the mesh shown in Fig.~\ref{fig:device}f to find these relations. The material parameters used for the Ip-Dip resist were $\rho=1100~\text{kg}/\text{m}^3$, $E=2.5~\text{GPa}$ and $\nu=0.49$ \cite{lemma2016mechanical}. A set of simulated lowest frequency relations for the shorter period PnC is shown in Fig.~\ref{fig:dispersion}, as a function of the absolute value of the phonon wavevector.  %For the film the emission power can be calculated in a similar way and the dispersion relations can be found easily using Lamb-mode theory \cite{graff1975wave,PhysRevB.70.125425}.

\begin{figure}
	\centering
    	\includegraphics[width=\columnwidth]{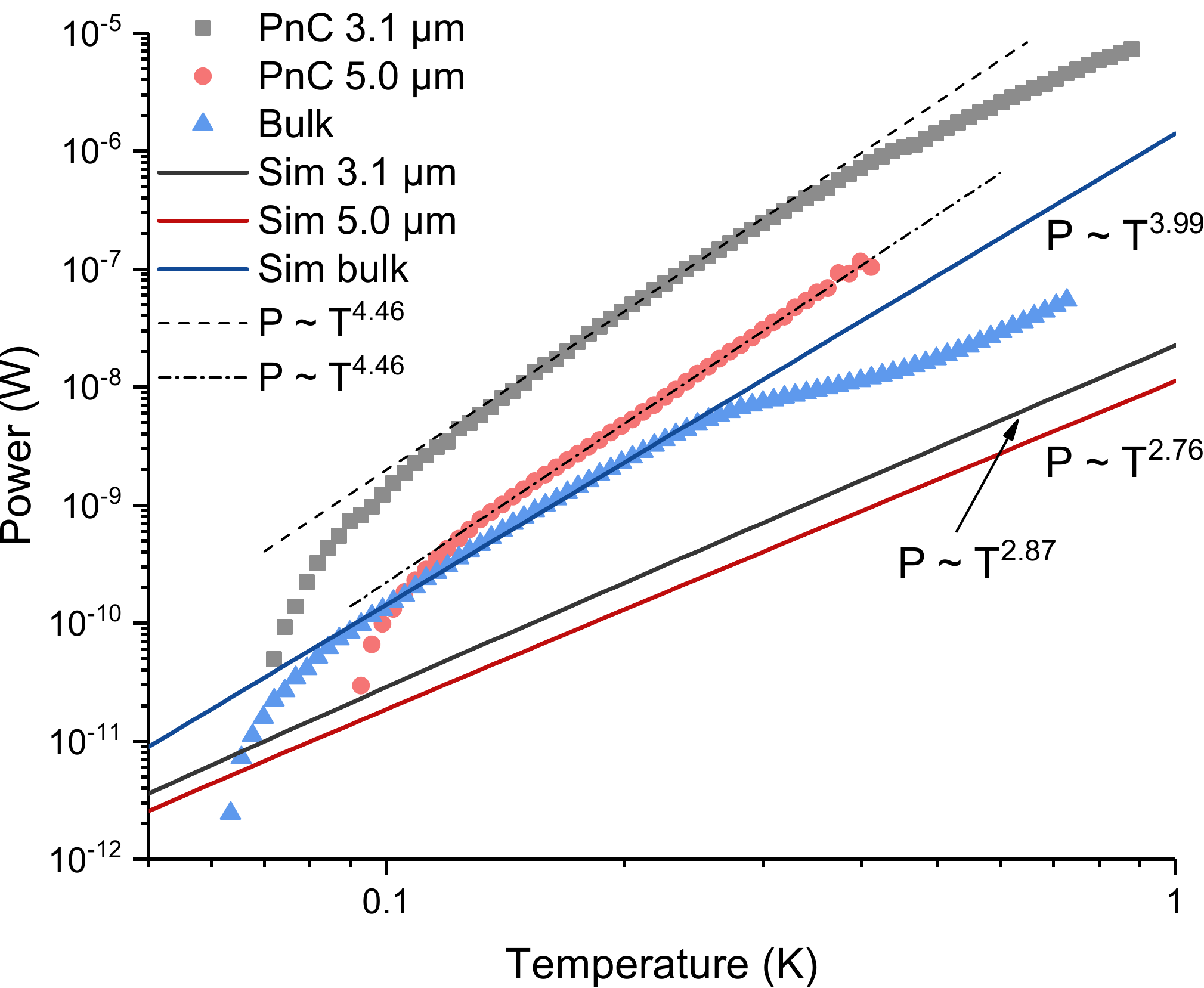}
    	\caption{The results of the thermal conductance measurements presented as the measured heating power of the SINIS heater (emitted phonon power) as a function of the measured temperature on the platform (colored circles). The colored lines represent the results given by the FEM simulations for each sample, without any fitting parameters, assuming zero bath temperature for clarity. The dip at the high temperature end of the bulk data is likely caused by a non-ideality in the behaviour of the thermometer junction.}
    	\label{fig:ballistic}
\end{figure}

The final simulated curves in Fig.~\ref{fig:ballistic} for the phonon emission power were calculated from the FEM simulated data for the platform and the PnC using the real physical dimensions of the metallic heater and the horizontal contact area between the platform and the PnC, or in the case of the bulk sample, the vertical contact area between then platform and the bulk support (smaller). A series thermal resistance model was used (platform + PnC or bulk support, Fig.\ref{fig:device}e), with the temperature at the interface %between the platform and the PnC or the bulk ("heater" of the PnC) was 
calculated by setting the power flowing through the platform and the PnC (bulk) to be equal. This modeling confirmed our earlier assertion that for the PnC samples, the coherent ballistic thermal conductance is expected fully limited by the PnC itself, as the interface and SINIS heater temperatures were very close to equal. %Due to the geometry these temperatures were very close to the temperature of the cuboid SINIS heater. Then using the interface temperatures the final power curves could be calculated.

\section{Results}

The thermal conductance of the fabricated structures was measured at sub-Kelvin temperatures utilizing a \ce{ ^{3}He}/\ce{ ^{4}He} dilution refrigerator. These measurements were made for the two different simple cubic PnC structures with lattice constants 3.1 $\upmu$m and 5.0 $\upmu$m, and also for a bulk sample with otherwise identical geometry, except that the PnC structure was substituted by bulk resist. The measurements were done by first calibrating one of the SINIS junction pairs to work as a thermometer, by measuring the response as a function of varying refrigerator (bath) temperature. The thermal conductance experiment was then performed by Joule heating the second SINIS device, and simultaneously measuring the temperature of the platform and the dissipated power from $P=IV$. The heating was done with a very slowly sweeping power to maintain quasi-equilibrium. The results of these measurements are shown in Fig.~\ref{fig:ballistic} together with the results from the FEM simulations.

From the measurements it is clear that the PnC structures did alter the thermal conductance of the material very strongly, as there is a large difference to the bulk result, and even between the two PnC structures. This also proves that the transport is mostly ballistic, as diffusive FEM calculation showed identical results with all three geometries (limited by the platform). Also, it is noticeable that the ballistic modeling explains the bulk results surprisingly well, without any fitting parameters. The observed temperature exponent of the bulk device is the expected $P \sim T^4$ for 3D ballistic conduction. 

However, the data for the PnC devices deviated very strongly from the modeling. As can be seen in Fig.~\ref{fig:ballistic} the coherent modeling predicts that the PnCs should decrease the thermal conductance, by about an order of magnitude, %via coherent modification of the phononic modes
whereas the measurement shows a puzzling {\em increase}, up to an order of magnitude for the smaller period.  %The fact that the simulated curve for the bulk structure fits the measured data well, suggests that at least in the bulk structure the transport is completely ballistic. 
In addition, the observed temperature exponent for the PnC devices, $P \sim T^{4.5}$ does not agree with the coherent modeling result, where a much weaker power law is expected. Very little is currently understood of this surprising enhancement. We speculate that it may have to do with thermal boundary resistance physics between a PnC and a solid structure, which has not been studied at all, and is therefore not understood yet.  

% However, excluding their respective order the two PnC structures differ largely from the model, which might be caused by a difference in Kapitza resistances. It is possible that there is a large difference in the Kapitza resistances since in the PnC structures there is a large interface between the film and the crystal. In the bulk structure instead, there is a direct transition from film to bulk at the edge of the suspended film leading to a very small contact area.

\section{Conclusions}

We have shown that 3D direct-laser-write two-photon lithography is a viable method for the fabrication of 3D phononic crystal structures for the hypersonic frequency range, which allows applications in thermal sciences. The strength of the method is not only in the fact that it can be used to produce complex 3D crystal structures; the same system can also be used for the fabrication of measurement devices on the crystals. The reported sub-Kelvin temperature range thermal conductance experiments showed large, up to an order of magnitude differences between the phononic crystal results and an identical bulk sample, demonstrating % The structures shown here can already be used
the potential for controlling thermal transport using 3D structures. % even though they only allow the enhancement of thermal conductance. 
However, the simplest FEM simulations based on the picture of fully coherent modification of the phononic band structure did not capture the essential features of the data. In particular, the used modeling predicts a reduction, whereas the experiments demonstrated a large enhancement of thermal conductance.
This surprising enhancement is not understood yet. %this is most likely due to the chosen geometry leading to a large difference in Kapitza resistances and it should be possible to use different geometries to reduce thermal conductance. The control shown here is already very strong, as the conductance was enhanced by more than an order of magnitude compared to the bulk. 

{\bf Acknowledgements}
This study was supported by the Academy of Finland project number 298667. Technical support by Zhuoran Geng is acknowledged.  
%\newpage


%\bibliography{reference}
%\bibliographystyle{plain}

\end{document}